\documentclass[runningheads, envcountsame, a4paper]{llncs}
\usepackage{amsmath}
\usepackage{amssymb}

\usepackage[pdftex]{graphicx}
\usepackage{epstopdf}
\usepackage{tikz}
\usetikzlibrary{positioning}
\usepackage{xspace}
\usepackage{xfrac}

\newcommand*{\bs}[1]{\mathbf{#1}}
\newcommand*{\E}{\ensuremath{\mathbb{E}}\xspace}
\newcommand*{\Prob}{\ensuremath{\mathbb{P}}\xspace}

\newcommand*{\mrm}[1]{\ensuremath{\mathrm{#1}}}
\renewcommand*{\mit}[1]{\ensuremath{\mathit{#1}}}
\renewcommand*{\mid}{\,|\,}
\newcommand*{\indep}{\ensuremath{\perp\!\!\!\perp}}
\newcommand*{\ie}{i.e.\@\xspace}
\newcommand*{\eg}{e.g.\@\xspace}
\newcommand*{\tit}[1]{\textit{#1}}
\newcommand*{\etal}{\textit{et al.}\xspace}

\newcommand*{\stmc}{\textit{Short-term Momentum Change}\xspace}
\newcommand*{\stmcShort}{STMC\xspace}
\newcommand*{\stmcVar}{\ensuremath{Y_t^\lambda}\xspace}
\newcommand*{\stmcA}{\ensuremath{\Delta P_t^{t+\lambda}}\xspace}
\newcommand*{\stmcB}{\ensuremath{\Delta P_{t-\lambda}^{t}}\xspace}
\newcommand*{\stmcBc}{\ensuremath{\Delta P_{t^\prime-\lambda}^{t^\prime}}\xspace}

\newcommand*{\OtherCovar}{\ensuremath{\bs{U}}\xspace}
\newcommand*{\ScorMar}{\ensuremath{P_t}\xspace}
\newcommand*{\Covar}{\ensuremath{\bs{X}_t}\xspace}
\newcommand*{\Period}{\ensuremath{Q_t}\xspace}
\newcommand*{\Seconds}{\ensuremath{S_t}\xspace}

\newcommand*{\TE}{\ensuremath{\mit{TE}}\xspace}
\newcommand*{\TEhome}{\ensuremath{\mit{TE}_h}\xspace}
\newcommand*{\TEaway}{\ensuremath{\mit{TE}_a}\xspace}

\usepackage{booktabs}
\usepackage[hyphens]{url}
\usepackage{microtype}
\usepackage[misc]{ifsym}

\begin{document}
\title{Stop the Clock: Are Timeout Effects Real?}
\toctitle{Stop the Clock: Are Timeout Effects Real?}
\author{Niander Assis [\Letter] \and
Renato Assun\c{c}\~{a}o \and
Pedro O.S. Vaz-de-Melo}
\authorrunning{N. Assis et al.}
\institute{Departmento de Ci\^{e}ncia da Computa\c{c}\~{a}o\\
Universidade Federal de Minas Gerais, Belo Horizonte, Brazil\\
\email{$\{$niander,assuncao,olmo$\}$@dcc.ufmg.br}}
\tocauthor{Niander~Assis, Renato~Assun\c{c}\~{a}o, and Pedro~O.S.~Vaz-de-Melo}
\maketitle              
\begin{abstract}
Timeout is a short interruption during games used to communicate a change in strategy, to give the players a rest or to stop a negative flow in the game. Whatever the reason, coaches expect an improvement in their team’s performance after a timeout. But how effective are these timeouts in doing so? The simple average of the differences between the scores before and after the timeouts has been used as evidence that there is an effect and that it is substantial. We claim that these statistical averages are not proper evidence and a more sound approach is needed. 
We applied a formal causal framework using a large dataset of official NBA play-by-play tables and drew our assumptions about the data generation process in a causal graph. 
Using different matching techniques to estimate the causal effect of timeouts, we concluded that timeouts have no effect on teams’ performances. Actually, since most timeouts are called when the opposing team is scoring more frequently, the moments that follow resemble an improvement in the team’s performance but are just the natural game tendency to return to its average state. This is another example of what statisticians call the \emph{regression to the mean} phenomenon.
\end{abstract}
\section{Introduction}
\label{sec:intro}
In sports, timeout is a short interruption in a play commonly used to stop a negative flow in the game, to discuss a strategy change, or to rest the players~\cite{Coffino}. As this is the most direct way coaches can intervene during a game, their influence and strategic ability is best expressed during these events. A timeout is usually called when a team has a rather long streak of score losses~\cite{Halldorsson2016,Zetou2008}.
Popular belief and research~\cite{gomez2011,Mace1992,permutt2011,Roane2004,Sampaio2013} have found a positive effect on teams’ performances after the timeout. That is, on average, the team asking for the timeout recovers from the losses by scoring positively immediately after.
This observed difference has been wrongly used as evidence that the timeout has a real and positive effect on teams’ performance. In order to answer such causal question, a formal counterfactual analysis should be used, and that is what we propose in this work.

There is an intense interest on causal models to analyze non-experimental data since causal reasoning can answer questions that machine learning itself cannot \cite{pearl2018book}. 
Our approach is built on top of these causal inference approaches that are briefly introduced in Section~\ref{sec:framework}.
For each timeout event at time $t_r$ in the database, we found a paired moment $t_c$ in the same game when no timeout has been called that serves as a \emph{control} moment for $t_r$, reflecting what would have happened the timeout had not been called.
This control moment is chosen based on other variables about the current game instant, which were drawn in a causal graph that depict our assumptions about the generation of the data and, as we will further discuss, asses if the causal effect can be estimated without bias.
In order to quantify how the game changed just after a given moment $t$,
we proposed \stmc (\stmcShort), which is discussed in Section~\ref{sec:stmc}. 

After using a \emph{matching} approach to construct our matched data (pairs of $(t_r, t_c)$), we found virtually no difference between the distribution of the \stmcShort for real timeouts and control instants, i.e., the estimated \emph{timeout effect} is very close to zero or non-existent.
Hence, we conclude that the apparent positive effect of timeouts is another example of the well-known \textit{regression to the mean fallacy}~\cite{barnett2004regression}. The dynamic match score fluctuates naturally and, after an intense increase, commonly returns towards a mild variation.
Thus, because timeouts are usually called near the extreme moments, as we will show, the game seems to benefit to those loosing. In summary, the main contributions of this paper are the following:

\begin{itemize}
\item We proposed a metric called \stmc (\stmcShort) to quantify how much the game momentum changes after a time moment $t_r$ associated with an event, such as a regular ball possession or an interruption of the game;
\item We collected and organized a large dataset covering all the play-by-play information for all National Basketball Association (NBA) games of the four regular seasons from 2015 to 2018. A single season has over 280 thousands \tit{game instants}, as we define in Section~\ref{sec:method}, and over 17 thousand timeout events;
\item After a detailed causal inference analysis to evaluate the timeout effect of the \stmc, we did not find evidence that the timeout effect exists or that its effect size is meaningful. Inspired by others, we did also consider two other settings in which the timeout effect could be different: (i) only the last five minutes of the games and (ii) everything but the last five minutes.
\end{itemize}

The next section describes the previous work carried out on the effect of timeouts and discusses how our work distance from them.
Section~\ref{sec:framework} gives a background on causal inference and the statistical models adopted.
We start Section~\ref{sec:method} by summarizing timeout rules in the NBA and describing our dataset, the play-by-play tables. In Section~\ref{sec:stmc} we introduce our outcome variable of interest, the \stmc, and in Section~\ref{sec:model-causal} our causal model. In Section~\ref{sec:data}, we describe our treatment and control groups and, in Section~\ref{sec:model-matching}, we explain our matching approaches. 
All the results are presented in Section~\ref{sec:results}. We close the paper in Section~\ref{sec:conclusion} with our conclusions.
\section{Related Work}
\label{sec:relwork}
Timeouts are used and implemented in team sports for several reasons, such as to rest or change players, to inspire morale, to discuss plays, or to change the game strategy~\cite{saavedra2012}. However, timeouts are mostly used to stop a negative flow in the game~\cite{Halldorsson2016,Zetou2008},
which is popularly referred as ``the game momentum.'' In basketball, momentum arises when one team is scoring significantly more than the other~\cite{Mace1992,Siva1992}. 

Several earlier studies analyzed the effect timeouts have for decreasing the opponent's momentum in the game~\cite{gomez2011,Mace1992,Roane2004,Sampaio2013}. These studies analyze the effect of timeouts on teams' performance just before and after it was called. For instance, by using a small sample of seven televised games from the 1989 National Collegiate Athletic Association (NCAA) tournament, Mace~\etal~\cite{Mace1992} recorded specific events of interest, which were classified as either \emph{Reinforcers} (\eg successful shots) or \emph{Adversities} (\eg turnovers) and verified that the rate of these events change significantly among teams in the 3 minutes before and after each timeout. They found that while the team that called the timeout improved its performance, the opponent team decreased it. Other works reached the same conclusions using similar methodologies and different data sets~\cite{gomez2011,Roane2004,Sampaio2013}.

To the best of our knowledge, Permutt~\cite{permutt2011} was the first to acknowledge the \textit{regression to the mean} phenomenon in such analysis. Permutt considered specific game moments---timeouts called for after a team suffered a loss of six consecutive points. Similar to others, the short-term scoring ratio was observed to be higher after timeouts. However, in contrast to others, the paper compares real timeouts with other similar game moments without a timeout. With such analysis, Permutt found that timeouts can be effective at enhancing performance, but at a small magnitude. The most significant result shows that the home-team with a ``first-half restriction'' presents a $0.21$ increase in average ratio for the next ten points. Calling a timeout predicts that the home-team will score $5.47$ out of the next ten points as opposed to $5.26$ points when a timeout is not called. Thus, the conclusion is that timeouts do not have any significant effect in changing the momentum of a game, \ie, using 6-0 runs as an indicator of instances where momentum would be a factor, teams were successful at ``reversing'' momentum even without the timeout as a mediator.

Although the work of Permutt~\cite{permutt2011} innovates by considering counterfactuals, the analyses still leave room for reasonable doubts about the reality of the timeout effect. It fails to take into account the existence of other important factors that could also influence on the momentum change and confound the true timeout effect. As a result, spurious correlations could have caused the lack of effect observed in the data. In our work, we take into account other factors such as coaches’ and team’s abilities, stadium and match conditions, clock time, quarter and relative score between the teams. More important, different from all the studies described in this section, we adopt a formally defined causal model approach~\cite{pearl2009causality} with a counterfactual analysis with its constructed control group. We show in a compelling way that timeouts do not have an effect in teams’ performances.
\section{The Causality Framework}
\label{sec:framework}
For illustration, consider $Y$ our outcome variable and $A \in \{0,1\}$ the \emph{treatment variable.}
Regardless of the actual value of $A$, we define $Y_{A=1}$ to be the value of $Y$ had $A$ been set to $A=1$ and $Y_{A=0}$ to be the value of $Y$ had $A$ been set to $A=0$.
We say there is a causal effect of $A$ on $Y$ if $Y_{A=0} \neq Y_{A=1}$ and, conversely, there is no causal effect or the effect is null if $Y_{A=0} = Y_{A=1}$.
These defined values are called \emph{potential outcomes}~\cite{neyman1923application}
because just one potential outcome is factual, truly observed, while the others are counterfactuals. Therefore, we cannot generally identify the causal effects of a single individual. This problem is known as the \textit{Fundamental Problem of Causal Inference (FPCI)}~\cite{holland1986statistics}.

Nevertheless, in most causal inference settings, the real interest is in the population level effect, or the average causal effect defined by $\E(Y_{A=1} - Y_{A=0})$. 
$\E(Y\mid A=1) - \E(Y\mid A=0)$ gives a reliable estimate in randomized experimental studies, where treatment $A$ is assigned randomly to each unit. However, in observational ones, we need to collect more information to control for and to make assumptions.
One important assumption is the \textit{conditional ignorability}~\cite{rosenbaum1983central}. This assumption is satisfied if, given a vector of covariates $\bs{X}$, the treatment variable $A$ is conditionally independent of the potential outcomes ($Y_{A=0}\indep A\mid \bs{X}$ and $Y_{A=1}\indep A\mid \bs{X}$), and there is a positive probability of receiving treatment for all values of $\bs{X}$ ($0 < \mrm{P}(A=1\mid \bs{X}=\bs{x}) < 1$ for all $\bs{x}$). The conditional ignorability assumption allows us to state that $\E(Y_{A=1} - Y_{A=0} \mid \bs{X}) = \E(Y\mid A=1, \bs{X}) - \E(Y\mid A=0, \bs{X})$ for every value of $\bs{X}$. This represents the basic rationale behind the \emph{matching} technique.

The simplest matching technique is the \emph{exact matching}. For each possible $\bs{X} = \bs{x}$, we form two subgroups: one composed by individuals that received the treatment and have $\bs{X}=\bs{x}$, and the other by individuals that did not receive the treatment and have $\bs{X}=\bs{x}$. 
Unfortunately, exact matching is not feasible when the number of covariates is large or some are continuous. As an alternative, examples are usually matched according to a distance metric $d_{ij} = d(\bs{x}_i, \bs{x}_j)$ between the covariate configurations of pairs $(i,j)$ of observations.
The \emph{Mahalanobis distance}~\cite{stuart2010matching} is a common choice for such distance metric as it takes into account the correlation between the different features in the vector $\bs{X}$.
Another option is to use \textit{propensity score}~\cite{rosenbaum1983central} for estimation of causal effects, which is defined as probability of receiving treatment given the covariates, \ie, $s(\bs{X}) = \Prob(A=1\mid \bs{X})$. Rosenbaum and Rubin proved that it is enough to just match on a distance calculated using the scalar scores $s(\bs{x})$, rather than the entire vector $\bs{x}$~\cite{rosenbaum1983central}.

In general, each matching approach can be implemented using algorithms that are mainly classified as either \emph{greedy} or \emph{optimal}. The \emph{greedy} ones, also known as \emph{nearest-neighbor matching}, matches the $i$-th treated example with the available control example $j$ that has the smallest distance $d_{ij}$. \emph{optimal matching}, however, takes into account the whole reservoir of examples since the goal is to generate a matched sample that minimizes the total sum of distances between the pairs. Such optimal approach can be preferred in situations where there are great competition for controls. For a good review on different matching methods for causal inference, see~\cite{stuart2010matching}.

Whatever matching technique used, its success can be partially judged by how balanced out are the covariates in the treatment and control groups. By pruning unmatched examples from the dataset, the control and treated groups of the remaining matched sample should have similar covariate distributions, when we say that matching achieves \emph{balance} of the covariates distribution.
\section{The Causal Effect of Timeout}
\label{sec:method}
According to the NBA 2016-2017 season official rules, in a professional NBA regular game, each team is entitled to six full-length timeouts and one 20-second timeout for each half. A full-length timeout can be of 60 seconds or 100 seconds, depending on when the timeout was requested. Also, every game has four regular periods plus the amount of overtime periods necessary on the occurrence of ties.
There is a specific amount of timeouts expected in each period for commercial purposes.
If neither team calls a timeout before a specific time, thus not fulfilling the next expected timeout, the official scorer stops the game and calls a timeout. The timeout is charged to the team that has not been charged before, starting with the \emph{home} team. These timeouts are called \tit{mandatory} or \tit{official} timeouts.

In basketball games, possessions are new opportunities to score in the game. Each possession starts from the moment a team gets hold of the ball until one of his players scores, commits a fault, or loses the ball in defensive rebounds or turnovers. The total number of possessions are guaranteed to be approximately the same for both teams at the end of a match, so it provides a good standardization for the points scored by each team~\cite{kubatko2007starting}. Indeed, most of basketball statistics are already given in a per possession manner.

Play-by-play tables capture the main play events such as goal attempts, rebounds, turnovers, faults, substitutions, timeouts and end of quarters (periods). For each play, we have the time in which it happened, the players and/or the team involved and any other relevant information, \eg, the score just after the play is recorded. Each play event is recorded as a new line in the table. While ball possessions are not clearly recorded in play-by-play tables, one can identify every change of possession from observing the game events.

In this work, we use play-by-play tables to identify the ball possessions and use them to observe how the teams' performance change when timeouts are called. We have identified every change of possession alongside the main interruptions---timeouts and end of quarters---in each game. Every change of possession and every main interruption is considered a new \textit{game instant}. We model each basketball game as a series of discrete \emph{game instants}. More formally, a \emph{game instant} is either (1) a regular ball possession; or (2) a major game interruption, which can be a regular timeout, an official scorer timeout, or the end of a quarter. Player substitutions and fouls were not considered as a main interruption. In fact, substitutions can happen when the ball stops and not only during timeouts.

\subsection{Short-term Momentum Change}
\label{sec:stmc}

Here we describe our outcome variable associated with teams' performance, for which we aim to estimate how it is affected by the timeouts.
Let \{$P_t$\} be an univariate stochastic process associated to a single match and indexed by the discrete game instants $t$. 
At any timeout moment, the team calling the timeout is defined as the \textit{target team}.
The $P_t$ random variable at the end of the $t$-th game instant is the score of the target team minus the opposing team's score and it is called the scoring margin. Hence, $P_t$ is a positive quantity when the target team is winning the match at game instant $t$ and negative otherwise. 
At the end of the $t$-th game instant, $P_t = P_{t-1}$ in two situations. First, if $t$ is a regular ball possession instant whose attacking team did not score, \ie, the possession ended with a turnover or defensive rebound. Second, if $t$ is a main interruption instant and, consequently, none of the teams had the opportunity to score.

In order to evaluate how ``momentum'' changes after a game instant, we use the \stmc (\stmcShort), which is the amount by which the \tit{scoring margin per possession rate} changes right after an game instant.
For any game instant $t$ and a positive integer $\lambda > 0$, we define the \stmcShort, $\stmcVar$, as the the average rate of change from $P_t$ to $P_{t+\lambda}$ (\stmcA) minus the average rate of change from $P_{t-\lambda-1}$ to $P_{t-1}$ (\stmcB). Note that we do not take into account the possible change in scoring margin caused in game instant $t$ (the change from $P_{t-1}$ to $P_t$):

\begin{equation}
\stmcVar = \frac{P_{t+\lambda} - P_{t}}{\lambda} -
\frac{P_{t-1} - P_{t-\lambda-1}}{\lambda} = \stmcA - \stmcB
\label{eq:stmc}
\end{equation}
for $t-\lambda \ge 0$ and $t+\lambda \le n$, where $n$ is the total number of game instants in a given game. 

The hyper-parameter $\lambda$ controls the time window used to evaluated how the game scoring dynamics changes around $t$. To balance out the offensive and defensive ball possessions, $\lambda$ must be an even integer. 
Also, the variable $\stmcVar$ should only be evaluated if the interval $[t-\lambda,\,t+\lambda]$ contains no game interruptions, with the possible exception of $t$. 
In a causal perspective, $\lambda$ represents our assumption for how many game instants that the interference (calling a timeout or not) at game instant $t$ can influence and is influenced by, in the short-term.

Let $A_t$ be the binary indicator that a timeout has been called at time $t$. We will denote $A_t=1$ if a team calls a timeout right before the game instant $t$ and $A_t=0$ if $t$ is a regular ball possession. If we find the set of covariates $\bs{X}$ that satisfy the conditional ignorability assumption, we can apply a matching technique and our average causal effect of interest, $\E(Y^\lambda_{A_t = 1} - Y^\lambda_{A_t = 0})$, can be estimated taking the difference in means from the matched treatment and control groups. The estimated \emph{timeout effect} \TE is defined as:

\begin{equation}
\TE = \E(\stmcVar \mid A_t = 1) - \E(\stmcVar \mid A_t = 0). \label{eq:toeffect}
\end{equation}

Every game is composed by two teams, the \tit{home} and the \tit{away} team. Because we want to estimate the causal effect of timeouts on the performance of the \emph{team that actually asked for it}, we decided to estimate the average causal effect of timeouts called by the \tit{home} teams (\TEhome) and the \tit{away} teams (\TEaway), separately. We proceed now to present our causal model which encodes our assumptions. 

\subsection{The Causal Model}
\label{sec:model-causal}

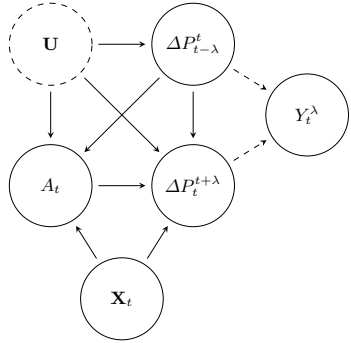
\begin{figure}[t]
	\centering
	\scalebox{0.75}{%
	\begin{tikzpicture}[%
		on grid=true,
		->,
		>=stealth,
		shorten <=3pt,
		shorten >=3pt,
		node distance=2.5cm,
		every node/.style={
		    circle,
		    minimum size=1.45cm,
		    draw}
		]
		\node (1) {$A_t$};
		\node[right=of 1] (2) {$\stmcA$};
		\node[above=of 2] (3) {$\stmcB$};
		\node[below right=2cm and 1.25cm of 1] (4) {$\Covar$};
		\node[dashed, above=of 1] (5) {$\OtherCovar$};
		\node[above right=1.25cm and 2cm of 2] (6) {$\stmcVar$};
		\draw (1) to (2);
		\draw (3) to (1);
		\draw (3) to (2);
		\draw (4) to (1);
		\draw (4) to (2);
		\draw (5) to (1);
		\draw (5) to (3);
		\draw (5) to (2);
		\draw[dashed] (2) to (6);
		\draw[dashed] (3) to (6);

	\end{tikzpicture}%
	}%
	\caption{A causal graph to model the timeout effect.}
	\label{fig:causal-graph}
\end{figure}

Pearl \cite{pearl2009causality} suggests the use of directed acyclic graphs (DAGs) as a way of encoding causal model assumptions with nodes representing the random variables and the direct edges representing direct causal relationships. One can identify in such graph a set of variables (or nodes) that satisfies the so called \emph{back-door criterion}~\cite{pearl2009causality}. These are variables that blocks all back-door paths from $A$ (the treatment variable) to $Y$ (the outcome variable) and does not include any descendants of $A$. Given that the graphical model includes all important confounding variables, it can be shown that conditioning on them suffices to remove all non-causal dependencies between $A$ and $Y$. In other words, it leaves only causal dependence that corresponds to the causal effect.

There are many factors that can potentially influence the short-term performance (\stmcShort) of the team that called a timeout after a given game instant. These can be \emph{intra-game} factors, which vary along the game, such as the scoring margin, the quarter and the time since the start of the quarter, or \emph{inter-game} factors, which vary from game to game, such as the venue conditions, the attendance at the venue, the specific adversary team, the players available and the teams' momentum in the season.

It is very intuitive why \emph{intra-game} factors, which are specific to a game instant, are considered a cause of both the treatment and outcome, thus being considered a confounder. In a not so straightforward way, some \emph{inter-game} factors are also very likely to affect both the treatment and outcome. For example, a team playing against a stronger or a weaker adversary would differently request the available timeouts and the afterwards performance may be differently affected.

Figure \ref{fig:causal-graph} shows our causal model graph. Each game instant $t$ can either receive the treatment assignment $A_t=1$ or $A_t=0$, meaning that the game instant is a timeout or a regular ball possession, respectively. The variables \Covar represent the observed covariates, which are \emph{intra-game} factors specific to the game instant $t$: (i) the current quarter (period) (\Period), (ii) the current scoring margin (\ScorMar) and (iii) the current time in seconds since the start of the period (\Seconds).
The variables represented by the node $\OtherCovar$ are the \emph{inter-game} factors, or the covariates related to a \textit{specific game} that influence both the treatment assignment and the game outcome as exemplified in the last paragraph.
Most of these variables are not directly observed or very difficult to measure---players and coach strategies, teams' relative skill difference and venue conditions. Hence, we include them in our graph as a dashed circle. The average rates of scoring margin change before ($\stmcB$) and after ($\stmcA$) the game instant $t$ are also in the graph, as well as the outcome $\stmcVar$ that is connected by dashed edges since it is a deterministic node---a logical function of the other two stochastic nodes.

We are interested in the causal effect of $A_t$ on \stmcVar. Since \stmcB is a direct cause of $A_t$ and not the reverse---for obvious chronological reasons---, we actually want to measure the causal effect of $A_t$ on \stmcA. 
According to the \emph{back-door} criterion~\cite{pearl2009causality}, if we adjust for $\OtherCovar$, $\Covar$ and $\stmcB$ we block any non-causal influence of $A_t$ on \stmcA. 

\subsection{Data}
\label{sec:data}
\label{sec:model-data}

Because we want to estimate \TE as defined in Equation~\eqref{eq:toeffect}, our treated and control groups are formed by game instants' \stmcShort, \stmcVar.
As discussed in Section~\ref{sec:stmc}, depending on which value we choose for $\lambda$, \stmcVar is not valid---if the short-term window induced by $\lambda$ includes another major interruption besides the possible $t$ or is longer than the start or end of the game. Therefore, our inclusion criteria for both groups is that \stmcShort exists and can be calculated.
For a given $\lambda$, the treated group, $\{\stmcVar | A_t = 1\}$, is formed by the valid real timeouts' \stmcShort. The control group, $\{\stmcVar | A_t = 0\}$, is formed by any valid game instant $t$'s \stmcShort that is not a timeout or any other kind of major interruption.

Since we want to estimate \TEhome and \TEaway, we have two treatment groups, one for timeouts called by \tit{home} teams and one for timeouts called by \tit{away} teams. On the other hand, it does not make sense to classify the control group as either \tit{home} or \tit{away}, thus we have just one control group. We will limit ourselves in the future to just mention theses treatment groups as either the \tit{home} treatment group or \tit{away} treatment group.

Our data consist of play-by-play information for every game from the 2014-2015, 2015-2016, 2016-2017 and 2017-2018 National Basketball Association NBA regular seasons. We crawled the data from the Basketball-Reference website\footnote{\url{http://www.basketball-reference.com}}. Most of our analysis will consist only of games from the NBA 2016-2017 season because using more than a single season would lead to very big samples that are impractical to apply our matching approaches. Also, while we did perform the same analysis using only other seasons, achieving very similar results, the choice for the 2016-2017 season is arbitrary, mainly due to be the first season for which we collected the data.

The 2016-2017 season had a total of 30 teams and 1,309 games (1,230 for the regular season and 79 in the playoffs). Considering all games, we computed 281,373 game instants, including the 17,765 identified timeouts (7,754 were called by \emph{home} teams and 8,011 by \emph{away} teams), and the 2,000 \emph{mandatory} timeouts. Our datasets, code and further instructions on how to reproduce our results can be found at our GitHub repository\footnote{\url{https://github.com/pkdd-paper/paper667}}.

\subsection{Matching}
\label{sec:model-matching}

The variables \OtherCovar, \Covar and \stmcB should be controlled for. In other words, they should be considered as possible confounders. Consequently, all of these variables are included in our matching for a valid causal inference. While we consider \OtherCovar, the \emph{inter-game factors}, unobserved covariates, we can still control them by pairing timeout examples with non-timeout examples \textit{taken from the same game}.
Furthermore, the variable \stmcB is likely the most important confounder covariate in our model. Indeed, coaches tend to call a timeout when their teams are suffering from a bad ``momentum'', evidencing great influence on the treatment assignment $A_t$. Also, \stmcA, the average rate of scoring margin change after a game instant $t$, should be highly causal dependent on \stmcB.
Therefore, in whatever matching approach we use, timeouts and control examples \textit{taken from the same game} and with equal \stmcB are going to be matched, hopefully, achieving balance for \Covar. We also restrict our matches to be constructed with \textit{non-overlapping ball possessions}. This restriction arises from our assumption that $\lambda$ defines a range of game instants that are dependent and influence $A_t$ as discussed in Section~\ref{sec:stmc}.

We applied three matching procedures: (1) \textit{no-balance matching}; (2) \textit{Mahalanobis matching}, and (3) \textit{propensity score matching}.
In the no-balance matching, each treatment example is paired with a valid control example that has the same \stmcB and is taken from the same game. We did not considered \Covar in this matching.
For the Mahalanobis matching, we applied the Mahalanobis distance using all covariates in \Covar, \ie, current quarter (\Period), current scoring margin (\ScorMar), and current clock time in seconds (\Seconds) since the start of the quarter. Finally, for the propensity score matching technique, we applied a simple euclidean distance match on the estimated scalar propensity score.

The true propensity score $s(\bs{X}) = \Prob(A=1\mid \bs{X})$ is unknown and must be estimated. Since estimating $\Prob(A=1\mid \bs{X})$ can be seen as a classification task, any a supervised classification model could be used. 
While logistic regression is the most common estimation procedure for propensity score, Lee~\etal~\cite{lee2010improving} showed that, in a non-linear dependence scenario, the use of machine learning models such as boosting regression trees to estimate the propensity score achieves better covariate balance in the matched sample. Indeed, our treatment assignment present a great non-linear dependence on its covariates. Take the clock time \Seconds, for example. As explained in Section~\ref{sec:data}, the timeout rules of NBA stimulate coaches to call a timeout just before a mandatory timeout would have been called by the official scorer. We use the boosting regression tree algorithm implemented in the \emph{gbm} R package \cite{gbmsoftware} to estimate the propensity score using \Covar.

Because we restrict timeout and non-timeout pairs to be taken from the same game, we have a very sparse matching problem. The \emph{rcbalance} R package~\cite{Pimentel2016} implementation of \emph{optimal matching} exploits such sparsity of treatment-control links to reduce computational time for larger problems. We use the optimal algorithm implemented in this package for all the aforementioned matching approaches. In addition, before applying any matching technique, we retained in the control subpopulation only those non-timeout game instants $t^\prime$ ($A_{t^\prime}=0$) for which the value \stmcBc is exactly equal to at least one \stmcB calculated to a real timeout instant $t$ ($A_t = 1$) in the same game. This improved the running performance even more.
\section{Experimental Results}
\label{sec:results}
\begin{figure}[t]
    \centering
    \includegraphics[width=0.7\linewidth]{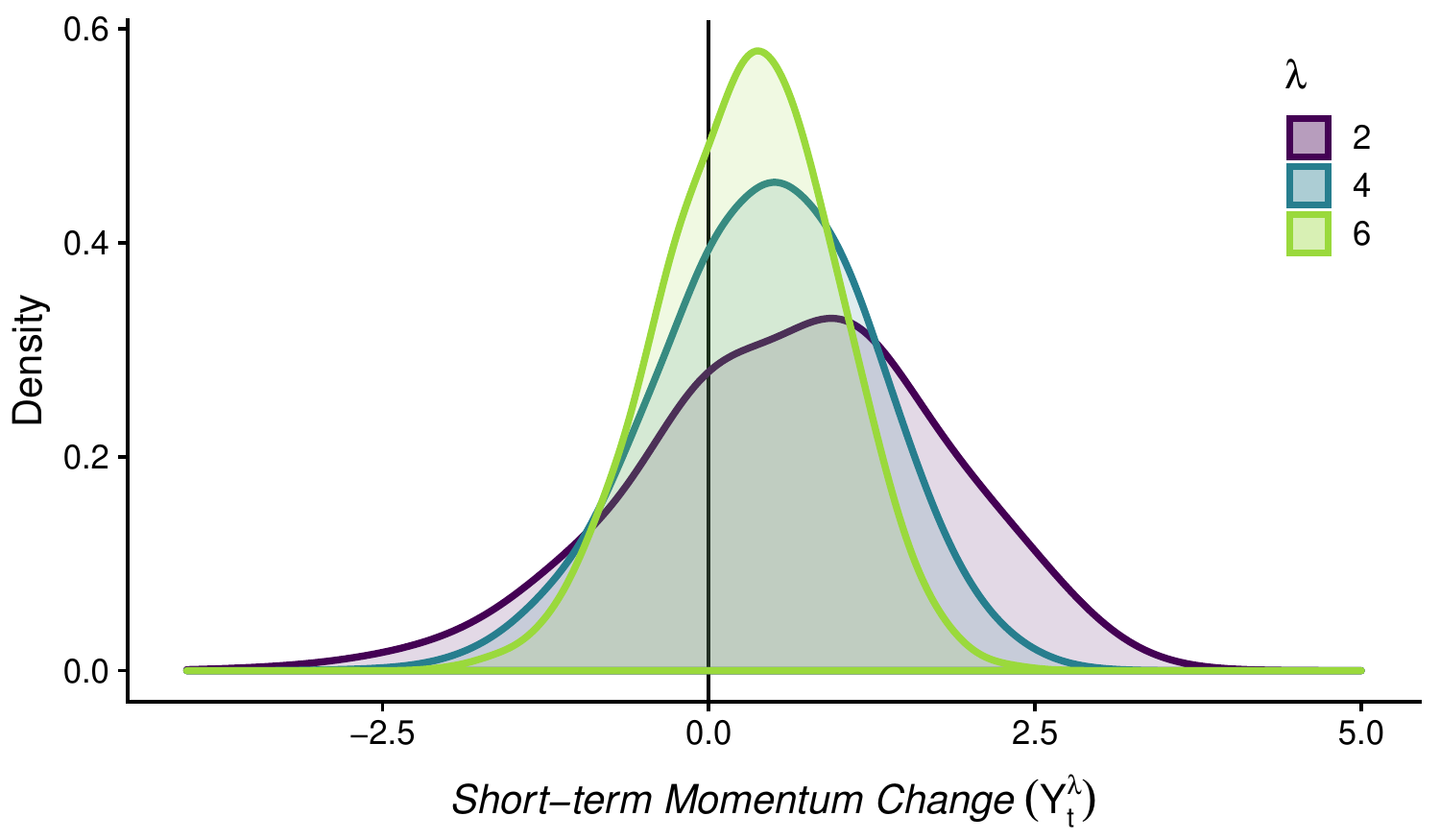}
    \caption{The \stmcShort ($\stmcVar$) distribution for home and away timeouts, considering three different ball possession windows $\lambda = 2,4,6$.}
    \label{fig:stmc-dist}
\end{figure}

In order to find out whether our data shows the generally accepted positive correlation between timeouts and improvements in the ``momentum'',
we calculated \stmcVar for every game instant associated with a timeout $t$ in every single game using $\lambda = 2,4,6$. 
Figure~\ref{fig:stmc-dist} shows the estimated density distribution of the \stmcShort \stmcVar for all timeouts, including those called by both \tit{home} and \tit{away} teams, but removing the \tit{official} timeouts.
The sample means and number of valid timeout examples in each sample are $0.629$ and $14,031$ for $\lambda=2$, $0.421$ and $12,225$ for $\lambda=4$, $0.302$ and $10,296$ for $\lambda=6$, respectively.

These results shows that, when a timeout is called by a team, its momentum improves by a small positive amount afterwards. For instance, with $\lambda = 4$, the average value of \stmcShort is $0.421$. That is, on average, 
there is an increase of $0.421$ points for a team's scoring margin per possession after it called the timeout. These results are consistent with previous works mentioned in Section~\ref{sec:relwork}~\cite{Mace1992,Roane2004,permutt2011,gomez2011,Sampaio2013}.
We applied the non-parametric ones-sample Wilcoxon statistical test and a bootstrap based test for the mean with the null hypothesis being that the mean is equal to zero.
Both tests for the three different values of $\lambda$ yielded p-values numerically equal to zero.

While these results could be used as evidence on why there is such common and widespread belief that timeouts improves teams' performance, or\tit{breaks} the momentum, it is not an evidence of the causal effect of timeouts. We move on to consider the analysis under our causal framework discussed in Section~\ref{sec:method}.

\subsection{Matching Results}

\begin{table}[t]
\centering
\caption{Summary statistics and SMD for balance assessment for matching using \tit{home} treatment group. The control ($A_t=0$) and timeout ($A_t=1$) groups are presented before (BM) and after all three matchings approaches: No-Balance (NB), Mahalanobis distance (M), and Propensity score (P)}
   \label{tab:tableone}
\resizebox{\textwidth}{!}{%
\begin{tabular}{ccccc@{\hskip 0.1in}ccc@{\hskip 0.1in}ccc} \toprule
&  & \multicolumn{3}{c}{\Seconds (mean(sd))}
       & \multicolumn{3}{c}{\Period (mean(sd))} 
       & \multicolumn{3}{c}{\ScorMar (mean(sd))} \\
    \cmidrule(lr){3-5}
    \cmidrule(lr){6-8}
    \cmidrule(lr){9-11}
    $\lambda$ & Method
        & $A_t=0$ & $A_t=1$ & SMD
        & $A_t=0$ & $A_t=1$ & SMD
        & $A_t=0$ & $A_t=1$ & SMD \\ \midrule
            2 & BM
    & 363.42 (198.77) & 410.03 (168.47) & 0.253
    & 2.42 (1.12) & 2.68 (1.15) & 0.222
    & 1.73 (10.81) & -0.00 (10.78) & 0.161 \\
                 & NB
    & 363.11 (201.26) & 410.20 (168.52) & 0.254
    & 2.47 (1.12) & 2.68 (1.15) & 0.178
    & 0.56 (10.63) & 0.00 (10.78) & 0.052 \\
                 & M 
    & 397.73 (168.83) & 410.27 (168.50) & 0.074
    & 2.61 (1.11) & 2.68 (1.15) & 0.062
    & 0.53 (10.86) & 0.00 (10.78) & 0.049 \\
                 & P 
    & 403.85 (163.08) & 410.14 (168.50) & 0.038
    & 2.67 (1.13) & 2.68 (1.15) & 0.009
    & 0.23 (11.01) & 0.00 (10.78) & 0.021 \\
         \midrule
                4 & BM 
    & 351.23 (187.75) & 388.45 (153.17) & 0.217
    & 2.33 (1.11) & 2.58 (1.13) & 0.222
    & 1.75 (10.69) & 0.22 (11.04) & 0.141 \\
                 & NB  
    & 351.59 (191.15) & 394.10 (151.23) & 0.247
    & 2.33 (1.11) & 2.58 (1.14) & 0.219
    & 0.44 (10.44) & 0.33 (11.14) & 0.011 \\
                 & M 
    & 381.35 (167.92) & 393.79 (151.12) & 0.078
    & 2.43 (1.04) & 2.57 (1.14) & 0.135
    & 0.50 (11.06) & 0.32 (11.12) & 0.016 \\
                 & P  
    & 385.03 (162.81) & 393.70 (151.37) & 0.055
    & 2.56 (1.11) & 2.58 (1.14) & 0.017
    & 0.38 (11.33) & 0.32 (11.12) & 0.005 \\
         \midrule
                6 & BM  
    & 334.96 (170.51) & 380.80 (143.08) & 0.291
    & 2.19 (1.08) & 2.49 (1.12) & 0.270
    & 1.67 (10.34) & 0.33 (11.09) & 0.124 \\
                 & NB 
    & 332.33 (173.87) & 389.79 (139.28) & 0.365
    & 2.20 (1.08) & 2.48 (1.12) & 0.253
    & 0.71 (10.40) & 0.79 (11.13) & 0.007 \\
                 & M 
    & 351.13 (162.90) & 389.67 (138.95) & 0.255
    & 2.27 (1.02) & 2.48 (1.12) & 0.200
    & 0.77 (10.80) & 0.80 (11.17) & 0.002 \\
                 & P 
    & 356.99 (160.61) & 389.72 (139.63) & 0.218
    & 2.35 (1.08) & 2.48 (1.12) & 0.122
    & 0.72 (10.91) & 0.81 (11.16) & 0.008 \\
           \bottomrule
\end{tabular}}%
\end{table}

Each of the three matching methods was applied twice:
one time using the treatment group with \tit{away} timeouts and the control group, and the other using the treatment group with \tit{home} timeouts and the control group. Some examples did not find a valid match and, therefore, were not included in the matched samples. Also, it should be noted that all matches were performed without replacement.

To evaluate for proper covariate balance between the treatment groups, a common numerical
discrepancy measurement is the difference in means divided by the pooled standard deviation of each covariate, known as the \textit{standardized mean difference} (SMD)~\cite{flury1986standard}. Unlike t-tests, SMD is not influenced by sample sizes and allows comparison between variables of different measured units. There is no general consensus on which value of SMD should denote an accepted imbalance level. Some researches, although, have proposed a threshold of 0.1~\cite{normand2001validating}. Table~\ref{tab:tableone} summarizes the covariate distribution with its mean and standard deviation values and the SMD of our matched samples considering all different approaches. 

For simplicity, we are only including here the results from the \tit{home} timeouts matched samples. Indeed, the \tit{away} samples showed very similar results and it can also be accessible from our GitHub repository\footnote{\url{https://github.com/pkdd-paper/paper667}}. 
Also, we do not show balance for $\stmcB$ as it is perfectly balanced due to our perfect match on this covariate. The unmatched sample sizes for control groups are $172,785$, $101,093$ and $49,403$; and, for treatment groups, $6,912$, $6,048$ and $5,127$ with $\lambda = 2,4,6$, respectively. For the matched samples, because of our 1:1 matching approach, the sample sizes are equal in both treatment and control groups, even across the different matching techniques---for $\lambda = 2,4,6$, they were $6,895$, $5,477$ and $3,832$, respectively.

From an initial look we can see that there are very similar results in terms of covariate balance for all three different matching approaches. Also, with no surprise, the matched treatment and control samples have equal sizes and the \stmcB covariate is equally distributed. Comparing the \emph{no-balance} matching with the balance in \emph{before matching}, it is clear that this simple matching procedure reduces substantially the SMD for all covariates. However, this is not enough to make the SMD negligible for all cases and hence whatever conclusions based on the comparison of these no-balance matched samples will be rightly subjected to doubt. Considering the \emph{Mahalanobis matching},
it achieved better covariate balance 
for all covariates and $\lambda$ values in comparison with \emph{before matching} and \emph{no-balance matching}. For $\lambda = 2$ and $\lambda = 4$, all SMDs are bellow $0.1$, with the exception of the \Period covariate in $\lambda=4$, in both analysis. The $\lambda=6$ configuration, on the other hand, presented the worse SMDs---\ScorMar is the only covariate with SMD bellow the $0.1$ mark.
The \emph{Mahalanobis matching} was not as good as the \emph{Propensity score matching}. All covariates have SMDs smaller than $0.1$ for $\lambda=2$ and $\lambda=4$ in both analysis. For $\lambda=6$, while \Period and \Seconds still had SMD above the $0.1$ mark, it is still smaller than the ones obtained in the previous matching for the same configuration.

\begin{figure}
    \centering
    \includegraphics[width=0.7\linewidth]{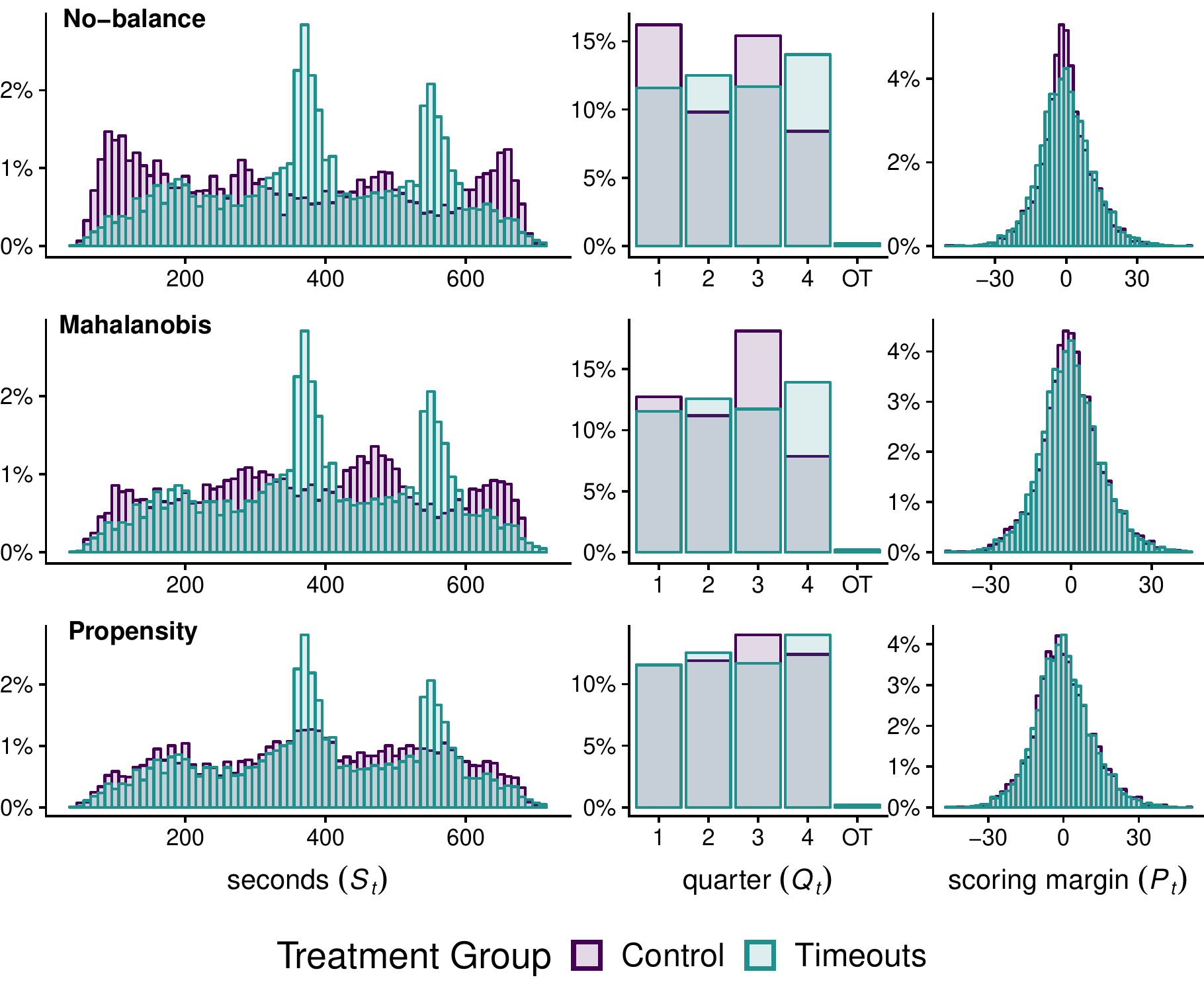}
    \caption{The control (magenta) and timeout (turquoise) \textit{seconds}, \textit{quarter} and \textit{scoring margin} covariate (\Covar) distributions after all matchings using the \tit{home} timeouts treatment group and $\lambda~=~4$.}
    \label{fig:balance-delta4-all}
\end{figure}

Plotting the covariate distributions from both treatment and control groups is a qualitative alternative of checking proper balance. Figure~\ref{fig:balance-delta4-all} shows the covariate distributions for \Seconds, \Period and \ScorMar after all three matchings with $\lambda=4$ on the analysis using timeouts called by the \tit{home} team. We can see that \emph{No-balance} and \emph{Mahalanobis} matching presented some imbalance for \Seconds and \Period. On the other hand, \emph{propensity score matching} shows rather similar distributions suggesting a much better balance. The fact that timeouts and control examples are matched only if taken from the same game makes it more complicated to find matches with a more balanced \Covar. \tit{Mahalanobis matching}, while trying to match samples as close as possible, encounters great difficulties. Propensity score, however, translates all interactions and non-linearities presented in the joint distribution of all covariates with the propensity score. This is why its matching was better.

\subsection{Timeout Effect}

We analyzed our matched data using a Monte Carlo permutation test~\cite[Vol II, chapter 10]{bickel2015mathematical} and difference of means between the two groups as our test statistic. When the null hypothesis is true, the timeout effect \TE defined in~\eqref{eq:toeffect} is equal to zero.
The treatment label was permuted ten thousand times.
It should be noted that we have large samples---the number of treatment-control pairs in each test varies from $3,766$ to $7,082$. 
Hence, we believe our analysis would mostly benefit if effect sizes, alongside statistical significance, are taken into account for our conclusions. Table~\ref{tab:results} shows the estimated average timeout effects for the \tit{home} (\TEhome) and the \tit{away} (\TEaway) teams. In addition, we have also included in the table a 99\% level confidence interval generated from the p-values obtained with the Monte Carlo permutation test.

There are two fundamental remarks here.
First, the estimated timeout effect \TE is very small for all cases, practically irrelevant during a game. 
Remember that the timeout effect \TE defined in Equation~\eqref{eq:toeffect} is the amount by which the team should expect its scoring margin per possession to change if a timeout were to be called by them. It would be rather challenging to find the minimum effect size for which it would be deemed enough to change the teams performance. It would not be a bad idea, however, to consider it to be an effect of at least \emph{one} marginal point. Specially because our assumption is that timeouts have an effect in the short-term window represented by $\lambda$. Yet, from our results, take the propensity matching with $\lambda=4$, which yielded the largest absolute timeout effect equal to $-0.059$, for instance. This means that if a team calls a timeout, it should expect its scoring margin to decrease $0.059$ points per possession in the next 4 possessions, which can be considered negligible in a basketball game.
Second, by analyzing the confidence intervals, the number of significant tests is very small given the very large number of examples in each of them. From the 18 tests, only 6 were statistically significant in the $\alpha=0.001$ level. In fact, these confidence intervals barely include the 0 effect value. Nevertheless, while these tests were statistically significant, they are still negligible effects. Also, we want to point it out that we did not perform any adjustment for multiplicity of tests. Indeed, such adjustment would yield a smaller number of statistically significant tests.

\begin{table}[t]
    \centering
    \caption{The estimated timeout effect for both \textit{away} (\TEaway) and \textit{home} (\TEhome) timeouts under different matching procedures and $\lambda$ values. The respective confidence intervals for each timeout effect for 99\% level is also shown. While some values are statistically significant, all timeout effects are negligible due to the small effect sizes.}
    \label{tab:results}
\resizebox{0.58\linewidth}{!}{%
\begin{tabular}{llcccc} \toprule
  $\lambda$ & Method & \TEaway & 99\% CI & \TEhome & 99\% CI \\ \midrule
  
    2    & No-balance & -0.028& (-0.075, 0.020)
& -0.022& (-0.072, 0.028) \\ 
    & Mahalanobis & -0.032& (-0.079, 0.015)
& -0.017& (-0.067, 0.032) \\ 
    & Propensity & -0.044& (-0.092, 0.004)
& -0.021& (-0.071, 0.028) \\ 
       \midrule
      4    & No-balance & -0.043& (-0.082, -0.004)
& -0.023& (-0.063, 0.017) \\ 
    & Mahalanobis & -0.053& (-0.092, -0.013)
& -0.013& (-0.053, 0.028) \\ 
    & Propensity & -0.059& (-0.098, -0.020)
& -0.032& (-0.072, 0.008) \\ 
       \midrule
      6    & No-balance & -0.031& (-0.068, 0.005)
& -0.036& (-0.072, 0.000) \\ 
    & Mahalanobis & -0.046& (-0.083, -0.010)
& -0.036& (-0.072, 0.000) \\ 
    & Propensity & -0.044& (-0.081, -0.008)
& -0.046& (-0.082, -0.009) \\ 
         \bottomrule
  \end{tabular}%
}
\end{table}

\begin{table}[t]
    \centering
    \caption{Additional analysis for excluding or considering only the last 5 minutes. For each analysis, the same matching approaches were applied using the new subsets of data. The results are very similar to the original analysis considering the whole game.}
    \label{tab:other-results}
\resizebox{\linewidth}{!}{%
\begin{tabular}{llcccccccc} \toprule
  & & \multicolumn{4}{c}{Minus Last 5 Min}
  & \multicolumn{4}{c}{Only Last 5 Min} \\
  \cmidrule(lr){3-6}
  \cmidrule(lr){7-10}
  $\lambda$ & Method & \TEaway & 99\% CI & \TEhome & 99\% CI
  & \TEaway & 99\% CI & \TEhome & 99\% CI \\ \midrule
    2    & No-balance & -0.029& (-0.082, 0.024)
& -0.010& (-0.065, 0.045)
& -0.083& (-0.115, -0.052)
 & -0.033& (-0.067, 0.001) \\ 
    & Mahalanobis & -0.021& (-0.074, 0.032)
& -0.010& (-0.065, 0.045)
& -0.091& (-0.122, -0.060)
& -0.040& (-0.074, -0.006) \\
    & Propensity & -0.044& (-0.098, 0.009)
& -0.012& (-0.067, 0.043)
& -0.094& (-0.125, -0.063)
& -0.038& (-0.072, -0.004) \\
       \midrule
      4    & No-balance & -0.036& (-0.078, 0.006)
& -0.030& (-0.073, 0.013)
& -0.006& (-0.068, 0.055)
& 0.038& (-0.029, 0.105) \\
    & Mahalanobis & -0.055& (-0.097, -0.013)
& -0.022& (-0.065, 0.020)
& -0.019& (-0.081, 0.044)
& 0.026& (-0.041, 0.093) \\ 
    & Propensity & -0.057& (-0.099, -0.015)
& -0.038& (-0.081, 0.004)
& -0.019& (-0.082, 0.043)
& 0.007& (-0.060, 0.074) \\ 
       \midrule
      6    & No-balance & -0.044& (-0.082, -0.005)
& -0.041& (-0.079, -0.003)
& -0.046& (-0.315, 0.223)
& -0.032& (-0.250, 0.186) \\
    & Mahalanobis & -0.051& (-0.089, -0.012)
& -0.040& (-0.078, -0.002)
& -0.028& (-0.299, 0.244)
& -0.048& (-0.267, 0.172) \\ 
    & Propensity & -0.046& (-0.085, -0.008)
& -0.055& (-0.093, -0.017)
& -0.046& (-0.317, 0.224)
& -0.048& (-0.265, 0.170) \\
         \bottomrule
  \end{tabular}%
 }%
\end{table}

We went further and investigated the \emph{timeout effect} \TE for two particular cases: (i) when the last five minutes are excluded and (ii) when only the final five minutes of the games are taken into account.
For both cases, we rerun the matching approaches on the new subsets of the data. Before executing the matching approaches, we filtered out from treatment and control groups examples that happened within the last five minutes of the last quarter (the 4th) of each game for (i). In a similar fashion, we filtered out from treatment and control groups examples that happened outside of the last five minutes of each game for (ii). However, because we ended up with fewer sample units available for matching, we included examples extracted from the other NBA seasons for the (ii) case, \ie, the 2014-2015, 2015-2016 and 2017-2018 seasons.

The results from both of these new analysis are in Table~\ref{tab:other-results}. 
For the case of excluding the last five minutes, we can see that we have slightly more statistical significant tests, 8 out of 18. Still, all of them have small effect sizes, making them not practically significant. The largest estimated effect is $-0.057$, which is found under the same configuration that we found the largest in the original analysis: \TEaway for the propensity score matching with $\lambda=4$. For the case of considering only the last five minutes, 
the only 5 statistically significant tests were all found with $\lambda=2$, but again, with very small effect sizes. The largest absolute effect is $-0.094$ for propensity score matching.
\section{Conclusion}
\label{sec:conclusion}
In this work we proposed a causality framework to quantify the effect of timeouts on basketball games. For the best of our knowledge, we were the first to resort on the theory of causality to solve this problem. While all previous studies pointed to a positive timeout effect, by applying our causality model on a large dataset of official NBA play-by-play data, we concluded that timeouts have no effect on teams' performance. This is another example of what statisticians call the \emph{regression to the mean} phenomenon. Since most timeouts are called when the opponent team is scoring more frequently, the moments that follow resemble an improvement in the team's performance, but are just the natural game tendency to return to its average state. We have also stratified our analysis by either including only the last five minutes or everything but the last five minutes of all games, but the results pointed to the same conclusion: timeouts have virtually no effect on team's performance.
%
%
\subsubsection*{Acknowledgments}
This work is supported by the authors' individual grants from FAPEMIG, CAPES and CNPq.
%
%

\end{document}